\documentclass[fleqn,twoside]{article}
\usepackage{espcrc2}

\usepackage{graphicx}
\usepackage{pstricks}
\usepackage{color}
\usepackage{units}
\usepackage{mcite}

\usepackage[figuresright]{rotating}

\newcommand{\AmS}{{\protect\the\textfont2
  A\kern-.1667em\lower.5ex\hbox{M}\kern-.125emS}}

\newcommand{\etal}{\MakeLowercase{\textit{et al.{}}}} 
\title{Influence of Low Energy Hadronic Interactions on Air-shower
Simulations}

\author{I.~C.~Mari\c{s}\address[KIT]{Karlsruhe Institute of Technology
    (KIT), D-76021 Karlsruhe, Germany}, R.~Engel$^{\rm a}$,
  X.~Garrido$^{\rm a}$, A.~Haungs$^{\rm a}$, M.~Roth$^{\rm a}$,
  R.~Ulrich$^{\rm a}$, M.~Unger$^{\rm a}$ }
\begin{document}

\begin{abstract}
  
  Experiments measuring cosmic rays above an energy of
  \unit[$\sim10^{14}$]{eV} deduce the energy and mass of the primary
  cosmic ray particles from air-shower simulations. We investigate the
  importance of hadronic interactions at low and high energies on the
  distributions of muons and electrons in showers on ground. In air
  shower simulation programs, hadronic interactions below an energy
  threshold in the range from \unit[80]{GeV} to \unit[500]{GeV} are
  simulated by low energy interaction models, like \textsc{Fluka} or
  \textsc{Gheisha}, and above that energy by high energy interaction
  models, e.g.\ \textsc{Sibyll} or \textsc{QGJSJet}. We find that the
  impact on shower development obtained by switching the transition
  energy from \unit[80]{GeV} to \unit[500]{GeV} is comparable to the
  difference obtained by switching between \textsc{Fluka} and
  \textsc{Gheisha}.

\vspace{1pc}
\end{abstract}

\maketitle

\section{Introduction}

Cosmic rays at very high energies ($E > \unit[10^{14}]{eV}$) can be
measured only indirectly by observing extensive air showers
in the atmosphere. Ground based air shower arrays in this energy range
observe the distribution of particles on observation level to deduce the mass and
the primary energy of the initiating particle. The relation between
the measurements and these quantities can be inferred only from
simulations of the development of the air
showers~\cite{Antoni:2005wq,*Knapp:2002vs}.

The present implementations of interaction models cannot describe all observed air shower
properties with a good precision. For example, at the Pierre Auger
Observatory, the observed number of muons is underestimated in
simulations with commonly used hadronic interaction
models~\cite{augerHDR}. 
The uncertainties in predicting shower properties are mainly
a consequence of uncertainties in the modeling of hadronic
interactions, since these cannot be obtained directly from QCD
calculations. 

Phenomenological hadronic interaction models are needed for describing particle
interaction properties over a wide range in energy, including phase
space regions and energies, which are presently not covered by
particle physics experiments. The energies of the first interaction in
high energy showers are not accessible by current accelerators. At low
energies there is a lack of precise data in the forward region, which
is relevant for air shower development~\cite{Engel:2002id}.

\section{NA61/SHINE Experiment}

Among the accelerator data results, the minimum bias analysis of p+p
and p+C~\cite{na49_2:2007,*na49_1:2006} collisions at a beam momentum
of \unit[158]{GeV/c}, delivered by the NA49 Collaboration, provided
information about the inclusive production of charged pions, which have
been already used to improve the precision of air shower
simulations~\cite{Ostapchenko:2005nj,EPOS}.  The NA61/SHINE apparatus is an upgrade of
the large acceptance experiment NA49~\cite{Afanasiev:na49}. A new time of flight
detector in the forward beam direction has been installed and tested,
increasing the accuracy of the particle identification. The update of
the DAQ and of the readout of tracking detectors provides an increase
of the maximum detection rate by a factor of 10 with respect to NA49.
In 2009 the data taking program for NA61/SHINE contains $\pi$+C
interactions at energies of 158 and \unit[350]{GeV}.

\section{Air-Shower Simulations}

\begin{figure*}[t]
\centerline{
  \includegraphics[width=0.93\columnwidth]{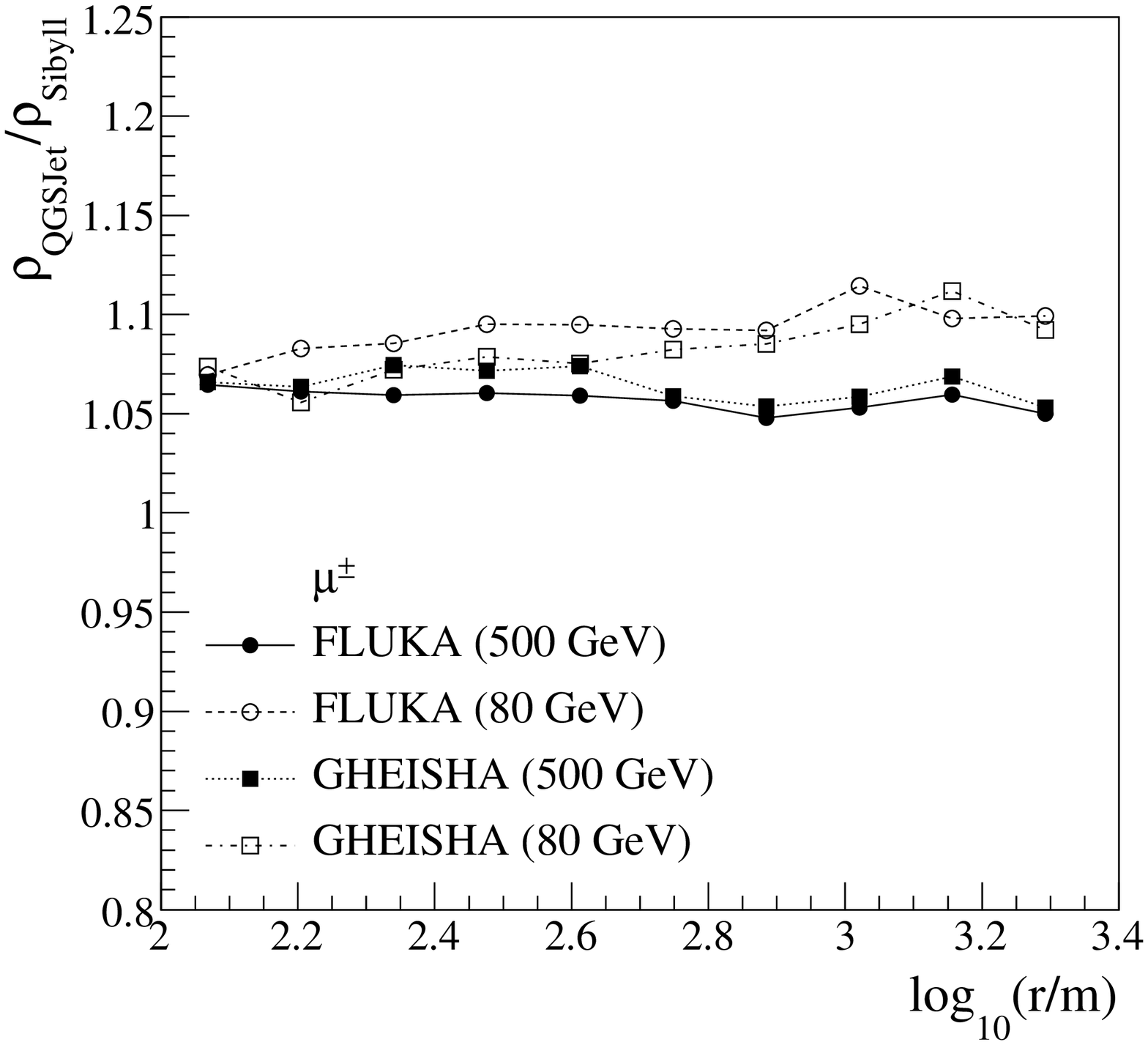}
  \includegraphics[width=0.93\columnwidth]{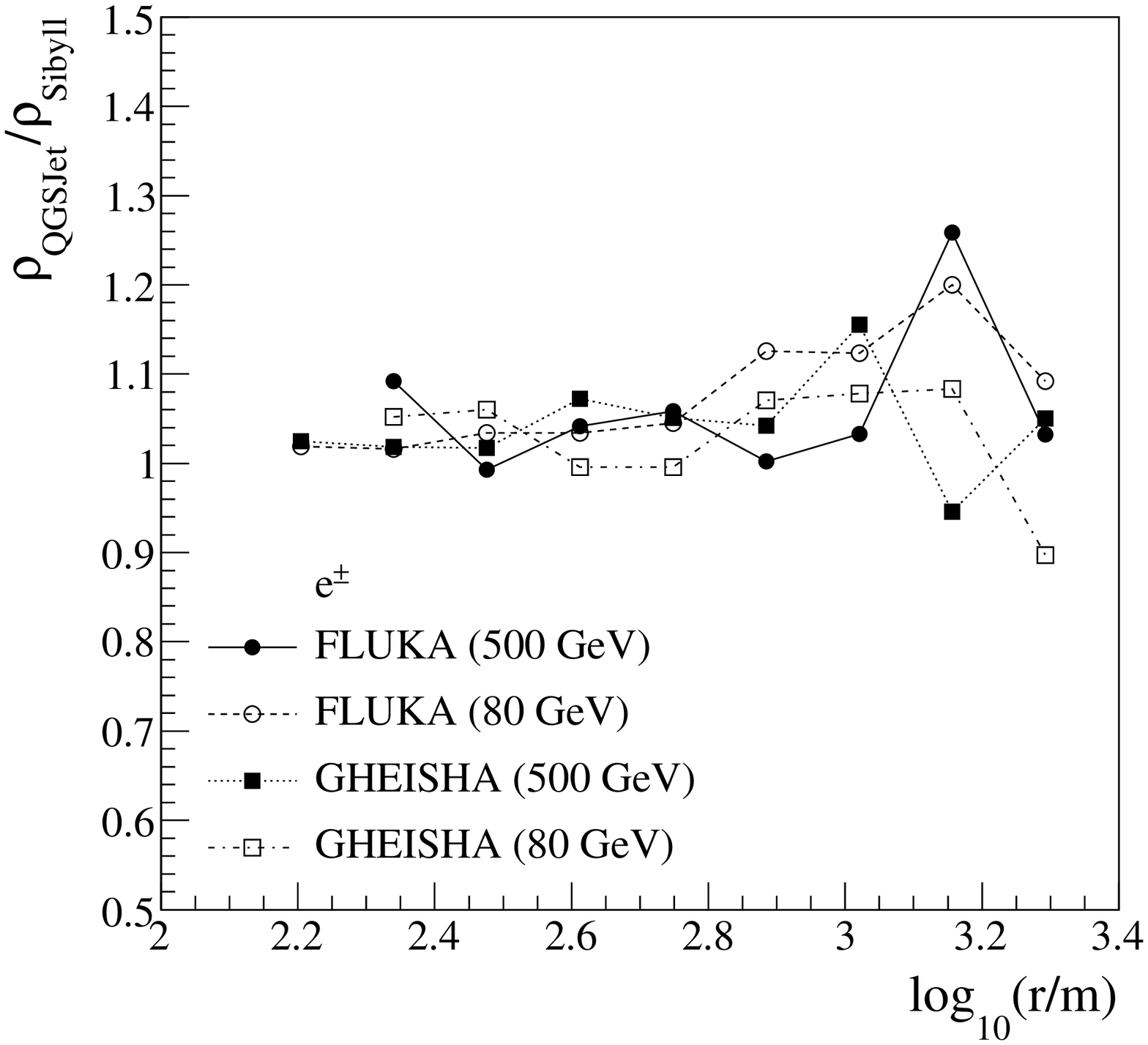}\\
}
  \vspace*{-1.1cm}
  \caption{Lateral distribution of the ratio of number of muons (left)
    and electrons (right) between \textsc{QGSJet} and \textsc{Sibyll}. The same hadronic
    interaction model is used for describing of the hadronic
    interactions below the transition energy indicated in brackets.}
  \label{fig:highmodel}
\end{figure*}

\begin{figure*}[t]
\centerline{
  \includegraphics[width=0.93\columnwidth]{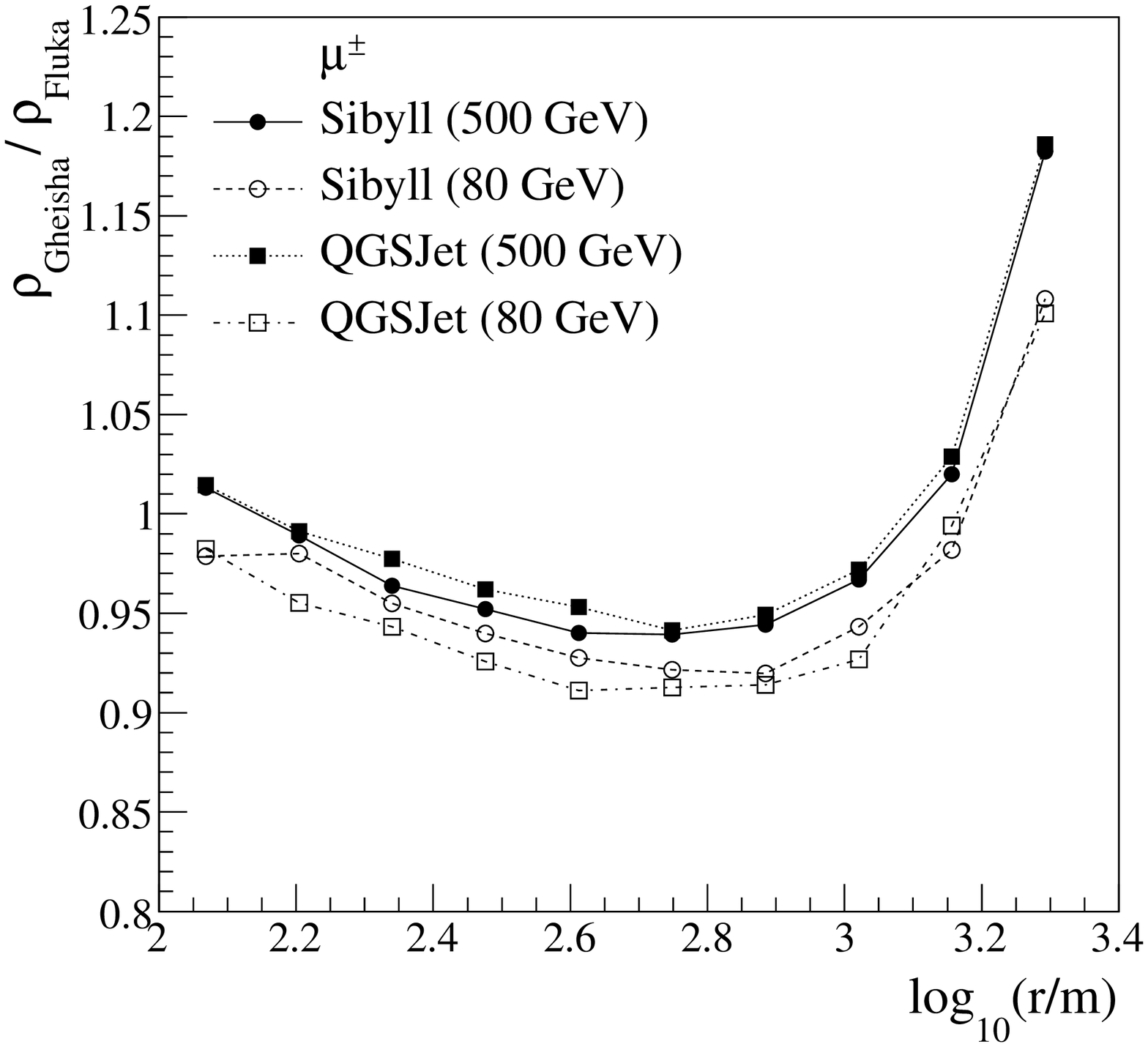}
  \includegraphics[width=0.93\columnwidth]{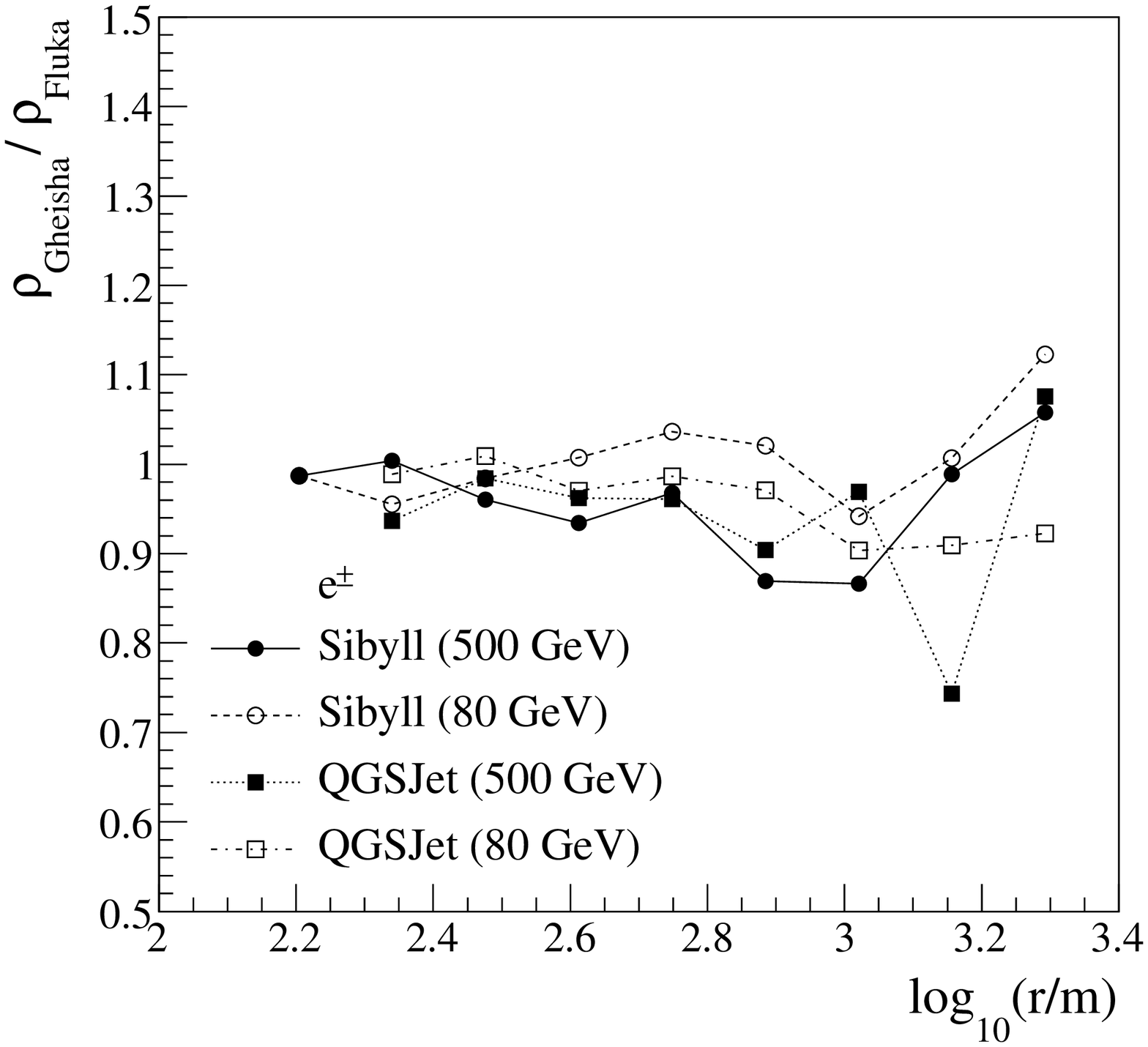}\\
}
  \vspace*{-1.1cm}
  \caption{Lateral distribution of the ratio of the number of muons
    (left) and electrons (right) as predicted by \textsc{Gheisha} and
    \textsc{Fluka}. The same hadronic interaction model is used for the
    description of the hadronic interactions above the transition
    energy indicated in brackets.}
  \label{fig:low}
\end{figure*}

\begin{figure*}[t]
\centerline{ 
  \includegraphics[width=0.93\columnwidth]{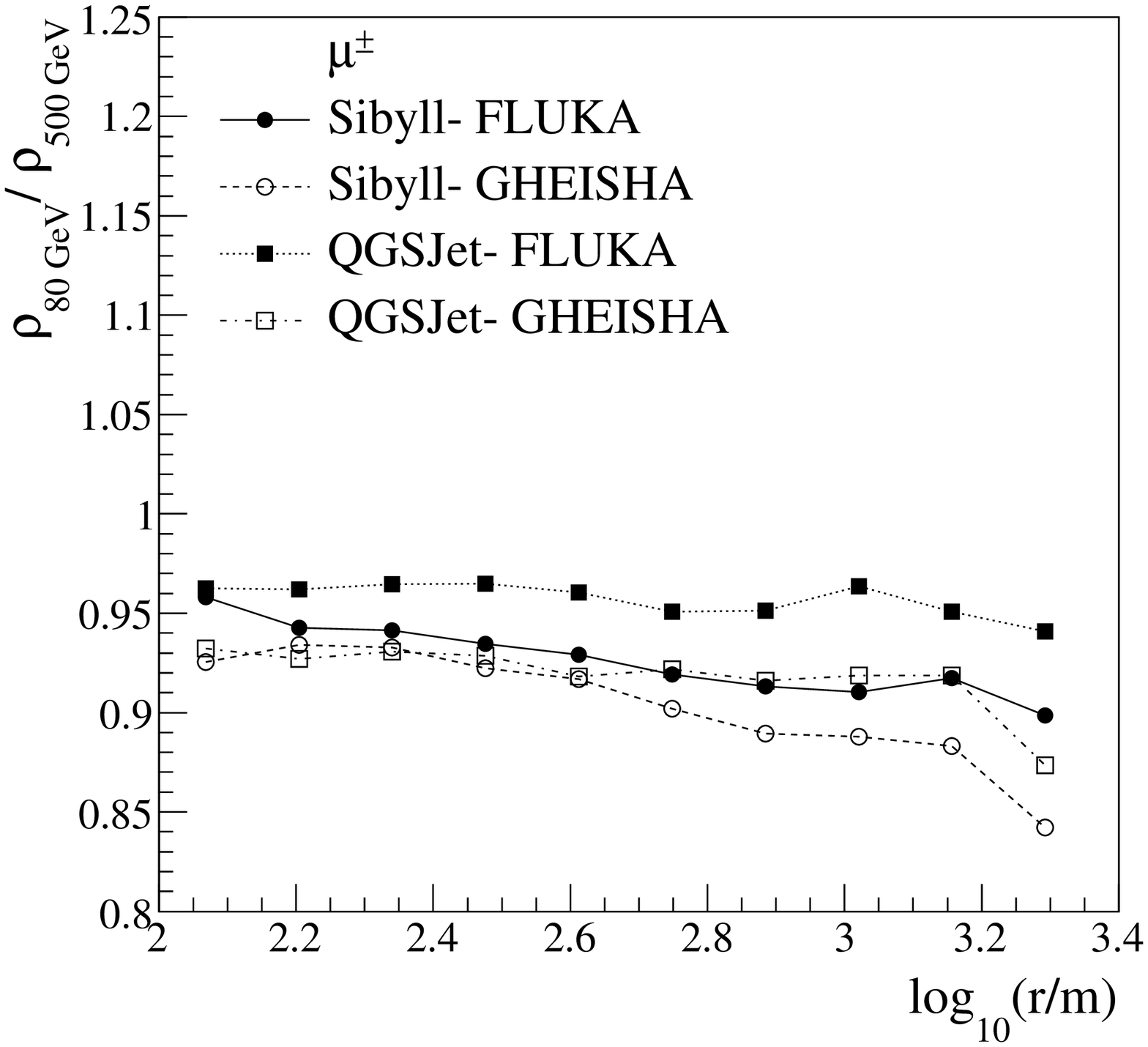}
  \includegraphics[width=0.93\columnwidth]{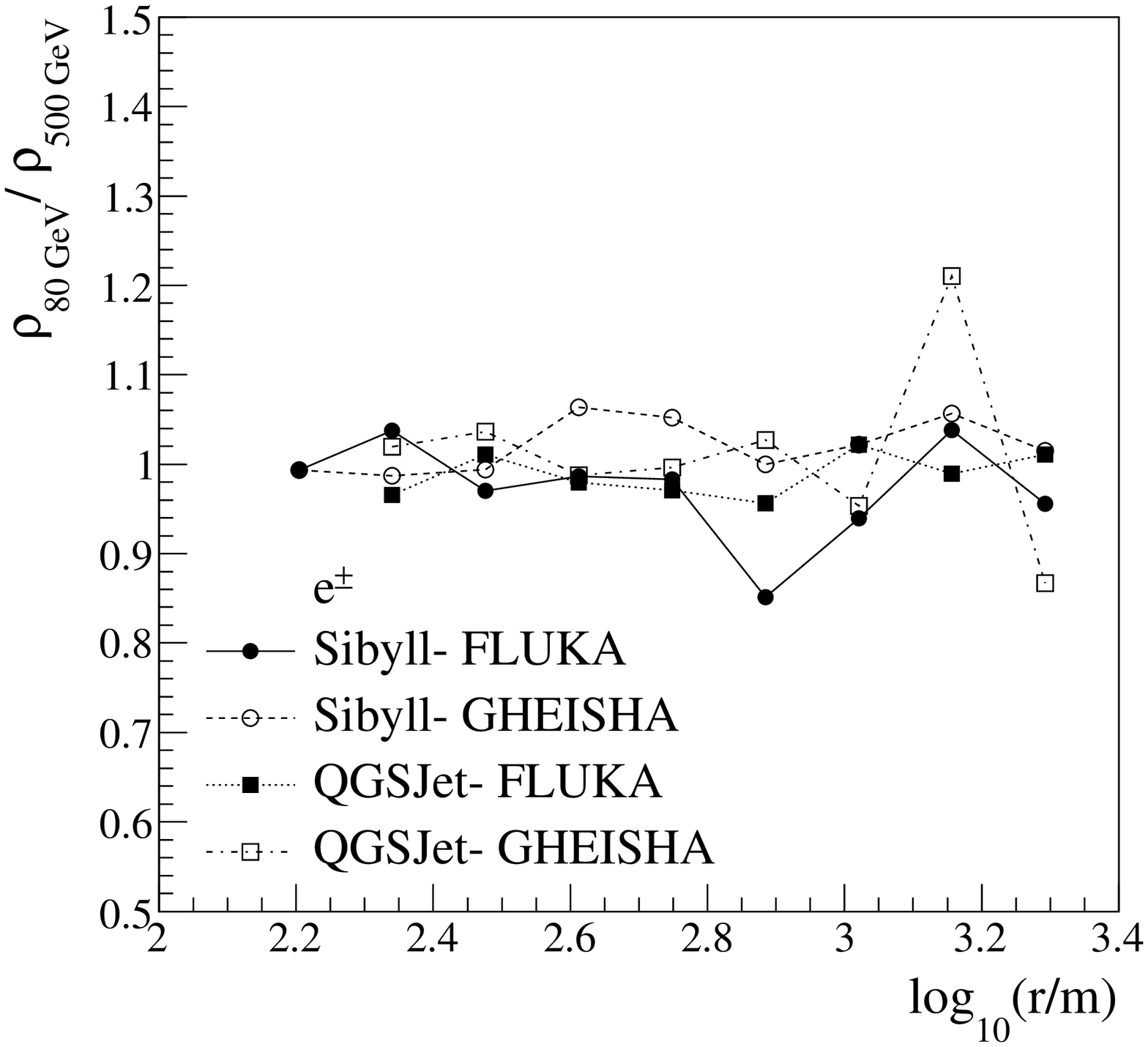}\\
}
  \vspace*{-1.1cm}
  \caption{Lateral distribution of the ratio of the predicted number of muons
    (left) and electrons (right) obtained by switching the transition
    energy from \unit[80]{GeV} to \unit[500]{GeV}. }
  \label{fig:trEnergy}
\end{figure*}

The particles from air showers arrive at the observation level extending over a large area
of several hundred meters to kilometers, with the maximum particle density near
the air shower axis. To predict the lateral distribution of particles
one needs to know the interaction cross section with the air nuclei,
and the properties of the final state produced in such an interaction,
namely the multiplicity, composition and momentum distribution of the
final state particles.  While the high energy interactions are of direct
relevance to the longitudinal shower development~\cite{Ulrich}, the
 particle production
at low energy is important for the lateral distribution of shower
particles at ground. The use of different hadronic interaction models
for low-energy interactions  leads to
significant differences of the predicted
particle distributions~\cite{Drescher1,*Drescher2,*Meurer:2005dt}.
The electromagnetic particles, originating from $\pi^{0}$ decays,
bremsstrahlung and $e^{\pm}$ pair production with a small component
from $\mu^{\pm}$ decays, are well described by QED. The muonic
component, produced mainly in decays of charged pions (95\,\%) and
kaons (3-4\,\%), depends upon the hadronic models used.

To investigate the importance  of hadronic interactions in the NA61/SHINE
beam energy region for ground-based shower observables we performed
simulations with different interaction models and varied the transition
energy between the low- and high-energy models. Using
CORSIKA~\cite{corsika}, eight sets of
proton showers of \unit[$3.16\times 10^{18}$]{eV} with a
uniform $\cos(\theta)$ distribution, $\theta$ being the zenith angle
ranging from 0  to 60 degrees, were generated. The sets
are built by combinations of generators describing the low energy hadronic
interaction, \textsc{Fluka}~\cite{Battistoni:2007zzb} and
\textsc{Gheisha}~\cite{geisha}, and generators describing the high energy hadronic
interactions, \textsc{QGSJet II}~\cite{Ostapchenko:2005nj} and
\textsc{Sibyll~2.1}~\cite{Fletcher:1994bd,*Engel:1999db}. The
transition energy between the
models is taken to be \unit[80]{GeV} or \unit[500]{GeV}.  
EGS4~\cite{egs4} is used for the simulation of electromagnetic interactions. 

To make
a direct comparison of the number of particles predicted with
different assumptions already with 100 showers, shower-to-shower fluctuations have to be
eliminated as far as possible. Most of the fluctuations in the
shower development, and thus in the number of particles, originate from the
fluctuations in the first few interactions. Therefore, the first interaction
in the simulated ensembles are preselected to have a secondary
particle multiplicity 
larger than 1000 and an inelasticity of at least 0.85. This is achieved 
by pre-simulating air showers with CONEX~\cite{conex} using \textsc{QGSJet~II}
and selecting showers with the  suited characteristics of the first
interaction. The secondary particles of these first interactions are then
fed to CORSIKA to generate the corresponding air showers.

The differences between \textsc{QGSJet} and \textsc{Sibyll} for the predicted number
of muons and electrons arriving at ground level as a function of the
logarithm of the distance to the shower axis are illustrated in
Fig.~\ref{fig:highmodel}. The range in lateral distance for the figures is chosen such
that the particle densities on ground are larger than
\unit[$10^{-3}$]{particles/m$^2$}.  \textsc{Sibyll} predicts by 5 to 10\% fewer
muons with respect to \textsc{QGSJet} over the whole distance to the
core. Moreover the ratio of the muon densities is not constant with
the distance to the shower axis. Switching the transition energy
between the models from \unit[80]{GeV} to \unit[500]{GeV} reduces the
difference by up to 4\%. The density of electrons, as
expected, is not influenced by the choice of the model.  The
difference on the muon/electron densities between the low energy
hadronic interactions is shown in Fig.~\ref{fig:low}. The large
dependence on the distance to the shower axis is mainly due to
different transverse momentum distributions predicted by the models.  At distances
larger than \unit[1000]{m}, which are of relevance to the Pierre Auger Observatory,
\textsc{Gheisha} predicts \unit[5-20]{\%} more muons than \textsc{Fluka}. The proton-air
and pion-air cross-sections for the two models are similar in the range
\unit[80]{GeV}to \unit[500]{GeV}, but still a 10\%
effect is visible just by switching the transition energy.

In Fig.~\ref{fig:trEnergy}, we show the ratios of the muon and electron
density distributions for the predictions obtained with
\unit[80]{GeV} and \unit[500]{GeV}  as transition energy.
The observed difference is a convolution of the
parametrization in the high energy and low energy models of the
hadronic interactions in the range of \unit[80-500]{GeV}.  The smallest
difference of about 4\%, constant over the whole distance range, is
observed in the case of the \textsc{QGSJet}-\textsc{Fluka} combination. Any combination
between \textsc{Gheisha} and other models gives a distance dependent
difference of up to 15\,\% at $\log_{10}(r/\unit[]{m})=3.4$.

\section{Conclusions}
The effect of the simulation of low energy hadronic interactions
on air shower development has been investigated by switching the
transition energy between the high- and low-energy hadronic interaction
models and by comparing different interaction models. The difference
in the predicted number of muons at ground level can be as large as
15\,\%, which is of the same order as the difference obtained by
changing the high-energy hadronic interaction model. Hence better
modeling of hadron production is needed not only at
the highest energies but also for energies up to \unit[500]{GeV}. The
NA61/SHINE data will cover a large region of the forward phase space
of low energy hadronic interactions with energies up to
\unit[350]{GeV} for improving the reliability of air shower simulations.

\end{document}